\documentclass{aip-cp}

\usepackage[numbers]{natbib}
\usepackage{rotating}
\usepackage{graphicx}
\begin{document}
	
	\title{Computation of NLO Processes Involving Heavy Quarks Using Loop-Tree Duality}
	
	\author{F\'{e}lix Driencourt-Mangin}
	
	\affil{Instituto de F\'{\i}sica Corpuscular, 
		Universitat de Val\`{e}ncia -- 
		Consejo Superior de Investigaciones Cient\'{\i}ficas,\newline
		Parc Cient\'{\i}fic, E-46980 Paterna, Valencia, Spain}
	\corresp{felix.dm@ific.uv.es}

\maketitle

\begin{abstract}
We present a new method to compute higher-order corrections to physical cross-sections, at Next-to-Leading Order and beyond. This method, based on the Loop Tree Duality, leads to locally integrable expressions in four dimensions. By introducing a physically motivated momentum mapping between the momenta involved in the real and the virtual contributions, infrared singularities naturally cancel at integrand level, without the need to introduce subtraction counter-terms. Ultraviolet singularities are dealt with by using dual representations of suitable counter-terms, with some subtleties regarding the self-energy contributions. As an example, we apply this method to compute the $1\to2$ decay rate in the context of a scalar toy model with massive particles.
\end{abstract}

\section{INTRODUCTION}
With the ever-improving quality of experimental data obtained at the LHC, developing theoretical tools to describe more and more complex physical processes has never been more relevant than it is today. Going to higher-orders in perturbation theory while being able to perform fast numerical computations is requested to achieve accurate theoretical predictions. The conventional way to compute Next-to-Leading Order (NLO) or Next-to-Next-to-Leading Order (NNLO) corrections is Dimensional Regularization (DREG) \cite{Bollini:1972ui,'tHooft:1972fi}. Within this formalism, real and virtual integrals are computed separately in $d=4-2\epsilon$ instead of $d=4$ dimensions, and singularities -- which manifest themselves under the form of $\epsilon$ poles -- are subtracted using one of the many existing techniques \cite{Kunszt:1992tn,Frixione:1995ms,Catani:1996jh,Catani:1996vz}, as guaranteed by the Kinoshita-Lee-Nauenberg (KLN) theorem \cite{Kinoshita:1962ur,Lee:1964is}. However, this approach may not be the most efficient when dealing with multi-loop multi-leg processes as one has to deal with complex final-state phase-spaces for each of the different contributions. Moreover, building all the necessary counter-terms can be far from being trivial and is in general very time-consuming.\\
We will present in this article an alternative method called Four-Dimensional Unsubtraction (FDU) \cite{Hernandez-Pinto:2015ysa,Sborlini:2016gbr,Sborlini:2016hat}, which is based on the Loop-Tree Duality (LTD) theorem \cite{Catani:2008xa,Bierenbaum:2010cy,Bierenbaum:2012th}. This method allows to combine, at the integrand level, virtual and real contributions using a suitable and physically motivated mapping between momenta, so the cancellation of singularities is achieved locally, meaning that the integration can be performed in $d=4$ dimensions.\\
\section{THE LOOP-TREE DUALITY}
The LTD theorem allows to rewrite loop scattering amplitudes as a sum of phase-space integrals, with slightly modified propagators. For a generic $N$-particle one-loop integral with massive scalar particles,
\begin{equation}
L^{(1)}(p_1,...,p_N)=\int_{\ell}\prod\limits_{i=1}^{N}G_F(q_i)=-2\pi\imath\int_{\textbf{q}}\sum{\rm Res}_{\{{\rm Im}(q_0)<0\}}\left[\prod\limits_{i=1}^{N}(G_F(q_i))\right]=-\sum\limits_{i=1}^{N}\int_{\ell}\tilde{\delta}(q_i)\prod\limits_{j=1,j\neq i}^{N}G_D(q_i;q_j)\,,
\end{equation}
where the internal momenta are written $q_i=l+p_1+...+p_i$, the Feynman propagators $G_F(q_i)=(q_i^2-m_i^2+\imath0)^{-1}$, and the so-called dual propagators $G_D(q_i;q_j)=(q_j^2-m_j^2-\imath0\,\eta\cdot k_{ji})^{-1}$. The delta function $\tilde{\delta}(q_i)=2\pi\imath\,\theta(q_{i,0})\,\delta(q_i^2-m_i^2)$ sets the internal lines on-shell by selecting the pole of the Feynman propagators with positive energy and negative imaginary part. Finally, $\eta$ is an arbitrary future-like vector with positive definite energy, and $k_{ji}=q_j-q_i$ defines the sign in front of the $\imath0$ prescription. This change of prescription, far from introducing additional difficulties, is essential for the consistency of the method. More details can be found in \cite{Catani:2008xa}.
\section{LOCAL CANCELLATION OF SINGULARITIES AT NLO}
In general, the procedure for computing NLO cross-sections is to consider and integrate separately real and virtual corrections. However, with the FDU technique, it is possible to perform the summation over degenerate soft and collinear states from both contributions before integration, thanks to the use of a suitable mapping of momenta between real and virtual kinematics. This is achievable because the infrared (IR) singularities are located in a compact region of the loop three-momentum, therefore the cancellations occur locally and the introduction of IR subtraction terms is not needed. For a process at one loop with $m$ final state particles, we can write
\begin{equation}
\label{SUMREALVIRTUAL}
\sigma^{{\rm NLO}}=\int_{m}d\sigma_{\rm V}^{(1,{\rm R})}+\int_{m+1}d\sigma_{\rm R}^{(1)}\,,
\end{equation}
where $d\sigma_{\rm V}^{(1,{\rm R})}$ denotes the renormalized virtual correction obtained using LTD and reads
\begin{equation}
d\sigma_{\rm V}^{(1,{\rm R})}=\sum\limits_{i=1}^{N}\int_{\ell}2\,{\rm Re}\,\big\langle\,\mathcal{M}_N^{(0)}\,|\,\mathcal{M}_N^{(1,{\rm R})}\,(\tilde{\delta}(q_i))\,\big\rangle\,\mathcal{O}_N(\{p_i\})\,, 
\end{equation}
where $\mathcal{M}_N^{(0)}$ denotes the $N$-leg scattering amplitude at leading order (LO), while $\mathcal{M}_N^{(1,{\rm R})}$ denotes the one-loop scattering amplitude, including the self-energy corrections of the external legs. The integral is then weighted by the measure function $\mathcal{O}_N$, used to define a given physical observable. Similarly, the real correction reads
\begin{equation}
\int_{m+1}d\sigma_{\rm R}^{(1)}=\sum\limits_{i=1}^N\int_{m+1}|\,\mathcal{M}_{N+1}^{(0)}(q_i,p_i)|^2\,\mathcal{R}_i(q_i,p_i)\,\mathcal{O}_{N+1}(\{p_j'\})\,,
\end{equation}
where the external four-momenta $p_j'$, the phase-space and the tree-level scattering amplitude have been rewritten in terms of the loop three-momentum and the external momenta $p_i$ of the Born process.\\

Although in DREG both ultraviolet (UV) and IR singularities are treated equally, in FDU they are considered separately\footnote{For instance, in the massless case, the self-energy contributions integrate to 0 in DREG and therefore are not taken into account in the traditional subtraction methods, because the UV and IR singularities cancel each other. However, they must still be considered within FDU.}. Indeed, it is necessary to discriminate them in order to build a complete LTD representation of the virtual contributions, as explained in \cite{Sborlini:2016gbr,Sborlini:2016hat}.\\
The UV singularities are dealt with by introducing counter-terms for the mass and the wave-function at the integrand level. Those counter-terms are calculated by expanding the unrenormalized amplitude around the UV propagator $G_F(q_{\rm UV})=(q_{\rm UV}^2-\mu_{\rm UV}^2+\imath0)^{-1}$, then choosing a renormalization scheme and finally adjusting subleading terms in order to subtract the right finite pieces. For instance, in the $\overline{\rm MS}$ scheme, this would be equivalent to make their integrated form in $d=4-2\epsilon$ exhibit only an $\epsilon$ pole, i.e. no finite part. The counter-terms are process-independent and only depend on the nature of the particles involved, so it is only necessary to compute them once. More details about the UV renormalization within FDU can be found in \cite{Sborlini:2016hat}.\\
IR singularities however, are more tricky to deal with. First, it is necessary to isolate each of them into different regions of the phase-space so that there is no more than one IR singularity of a given type -- soft or (quasi-)collinear -- in a given region. Then, for each region, one must apply to the real integrand a mapping that will make similar singular behaviors, for the virtual and the real contributions, match and therefore cancel each other locally. Finally, both contributions can be added, as shown in Equation (\ref{SUMREALVIRTUAL}), and the integration can be safely performed in 4 dimensions.\\

In the next section, we will apply the FDU technique to the three-point scalar function, and illustrate with a few expressions.
\section{APPLICATION OF THE FOUR DIMENSIONAL UNSUBTRACTION}
Let's consider a $1\to2$ process at NLO with one massless internal state and the remaining internal and outgoing particles with mass equal to $M$. To simplify the expressions, we define $m=\frac{2M}{\sqrt{s_{12}}}$ and $\beta=\sqrt{1-m^2}$. The final-state on-shell momenta are labeled as $p_1$ and $p_2$, with $p_1^2=p_2^2=M^2$ and with $p_1^{\mu}=\frac{\sqrt{s_{12}}}{2}(1,\mathbf{0},\beta)$ and $p_2^{\mu}=\frac{\sqrt{s_{12}}}{2}(1,\mathbf{0},-\beta)$. The incoming one is $p_3=p_1+p_2$ with virtuality $p_3^2=s_{12}>0$. We write the three internal momenta $q_i=\ell+\sum_{j=1}^ip_j$, and when they are on-shell, $q_i^{\mu}=\frac{\sqrt{s_{12}}}{2}(\xi_{i,0},~2\xi_i\sqrt{v_i(1-v_i)}\,\mathbf{e}_{i,\bot},~\xi_i(1-2v_i))$, with $\xi_{i,0}=\sqrt{\xi_i^2+m_i^2}$, $m_1=0$ and $m_2=m_3=m$.\\
Since the virtual decay-rate is given by
\begin{equation}
\label{virtualdecayrate}
\Gamma_{{\rm V}}^{(1)} = \frac{1}{2\sqrt{s_{12}}} \, \int d\Phi_{1\to 2} \, 2 \, {\rm Re} \, \big\langle \mathcal{M}^{(0)} | \mathcal{M}^{(1)} \big\rangle= - \Gamma^{(0)} \,  2\, g^2 \, s_{12}\, {\rm Re}[L^{(1)}_{m > 0}(p_1,p_2,-p_3)]\,,
\end{equation}
we apply the LTD to $L_{m>0}^{(1)}(p_1,p_2,-p_3)$. Its dual representation then consists in the sum of the 3 following integrals:
\begin{eqnarray}
I_1 &=& \frac{4}{s_{12}} \int \, \frac{\xi_{1,0}^{-1} \, d[\xi_{1,0}] \, d[v_1]}{1-(1-2 v_1)^2 \beta^2}~, \nonumber \\ 
I_2 &=& \frac{2}{s_{12}} \int  \frac{\xi_{2}^2 \, d[\xi_{2}] \, d[v_2]}{\xi_{2,0} \, 
	\left(1-\xi_{2,0}+\imath 0 \right) \left(\xi_{2,0}+\beta \, \xi_{2} \, (1-2 v_2)-m^2\right)}~, \\ 
I_3 &=&  -  \frac{2}{s_{12}} \int \frac{\xi_{3}^2 \, d[\xi_{3}] \, d[v_3]}{\xi_{3,0} \, \left(1+\xi_{3,0} \right) 
	\left(\xi_{3,0}-\beta \, \xi_{3} (1-2 v_3)+m^2\right)}\,,\nonumber 
\end{eqnarray}
with $d[\xi_i] = \frac{(4\pi)^{\epsilon-2}}{\Gamma(1-\epsilon)} \left(\frac{s_{12}}{\mu^2}\right)^{-\epsilon} \, \xi_i^{-2\epsilon} \, d\xi_i~$ and $d[v_i] = (v_i(1-v_i))^{-\epsilon} \, dv_i$. On the other hand, the real contributions are given by the interference terms, and are written as
\begin{equation}
\label{realdecayrate}
\widetilde{\Gamma}^{(1)}_{{\rm R},i} = \frac{1}{2 \sqrt{s_{12}}} \, \int\, d\Phi_{1\to 3} \, 2 \, {\rm Re} \big \langle \mathcal{M}^{(0)}_{2r} | \mathcal{M}^{(0)}_{1r} \big \rangle \, 
\mathcal{R}_i\left(y'_{ir} < y'_{jr}\right) \, , \qquad i,j=\{1,2\}~,
\end{equation} with $y'_{ir}$ being the normalized scalar product between $p'_i$ and the radiated particle four-momentum $p'_r$. In Equation~(\ref{realdecayrate}), we have applied the phase-space partition mentioned at the end of the previous Section. Explicitly, we define $\mathcal{R}_i=\{y'_{ir}<{\rm min}~y'_{jk}\}$ for $i\in\{1,2\}$. Then, for each region, we introduce a suitable mapping between real and virtual kinematics. For instance, in region~1 we use
\begin{equation}
p_r'^\mu = q_1^\mu\,,~~~~~p_1'^\mu=(1-\alpha_1) \, \hat p_1^\mu + (1-\gamma_1) \, \hat p_2^\mu - q_1^\mu\,,~~~~~p_2'^\mu=\alpha_1 \, \hat p_1^\mu + \gamma_1 \, \hat p_2^\mu\,,
\end{equation}
with $\hat p_1^\mu$ and $\hat p_2^\mu$ fulfilling $p_1^{\mu}=\frac{1+\beta}{2}\,\hat p_1^{\mu}+\frac{1-\beta}{2}\,\hat p_2^{\mu},~p_2^{\mu}=\frac{1-\beta}{2}\,\hat p_1^{\mu}+\frac{1+\beta}{2}\,\hat p_2^{\mu}$. The parameters $\alpha_1$ and $\gamma_1$ are computed by solving the on-shell conditions $(p_1')^2=M^2,~(p_2')^2=M^2$, their expressions can be found in \cite{Sborlini:2016hat}. A similar transformation is used for region~2. It is worth emphasizing that these transformations fulfill two roles: they move the IR singularities to the same integration points in the dual domain, which allows them to cancel each other locally, and they are optimized to deal smoothly with the massless limit in each region. Finally, after rewriting the real contributions in terms of the loop variables, we obtain
\begin{eqnarray}
\widetilde{\Gamma}_{{\rm R},1}^{(1)} &=& \Gamma^{(0)} \, \frac{2 a}{\beta} \, \int \, d\xi_{1,0} \, dv_1 \, 
\frac{ {\cal R}_1(\xi_{1,0}, v_1) \,  {\cal J}_1(\xi_{1,0}, v_1) \,  (1-\xi_{1,0} (1-v_1))^2}{\xi_{1,0}^2 \, (v_1+\alpha_1(1-2 v_1)) ((1-v_1)(1-\xi_{1,0})-\alpha_1(1-2v_1))}~,
\label{GammaTilde1TOY} \\ 
\widetilde{\Gamma}_{{\rm R},2}^{(1)} &=& \Gamma^{(0)} \, \frac{2 a}{\beta} \, \int \, d\xi_{2} \, dv_2  \, 
\frac{ {\cal R}_2(\xi_2,v_2)\, {\cal J}_2(\xi_2,v_2) \, (2+(1-2v_2) \, \xi_2 - \xi_{2,0})}{(1-\xi_{2,0}+\imath0) (\xi_{2,0} + (1-2 v_2)\, (1-2 \alpha_2) \, \xi_2 - m^2)}\,,
\label{GammaTilde2TOY} 
\end{eqnarray}
with $\mathcal{J}_1$ and $\mathcal{J}_2$ being the Jacobians of the two transformations, $\mathcal{R}_1$ and $\mathcal{R}_2$ restricting the integration domain associated to their respective region, and $a=g^2_S/(4\pi)$.
\begin{figure}[ht]
\includegraphics[width=0.5\textwidth]{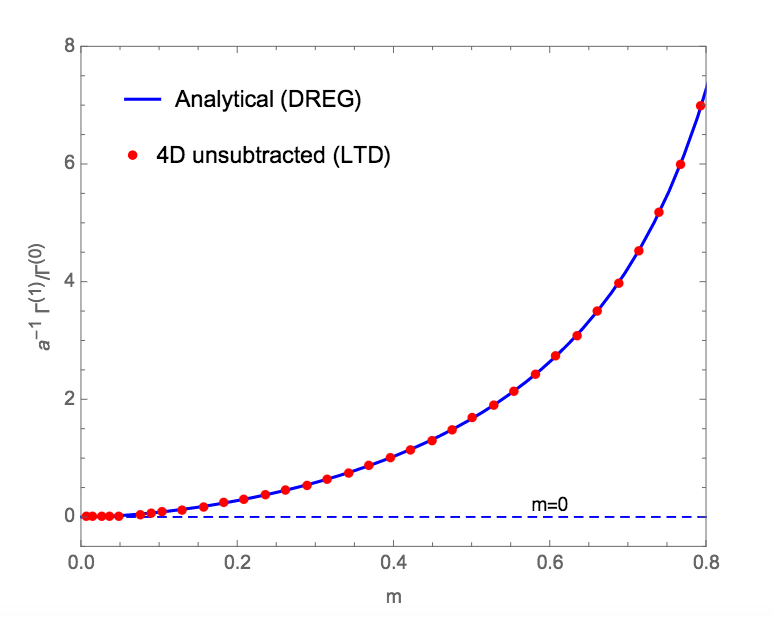}
\caption{Total normalized decay rate at NLO for a $1\to2$ process with scalar particles, as a function of the dimensionless parameter $m$. The solid line corresponds to the DREG analytical result, while the dots are the numerical values obtained using FDU. The horizontal line represents the massless limit.}
\label{plot}
\end{figure}
\newline
The sum of real and virtual contributions, i.e. the sum of Equations~(\ref{virtualdecayrate}), (\ref{GammaTilde1TOY}) and (\ref{GammaTilde2TOY}) , gives the total decay-rate at NLO, $\Gamma^{(1)}$, and is safely integrable in $d=4$ dimensions. The integration is done numerically, and is shown in Figure \ref{plot}. Our results are compared with the analytical expression obtained using the DREG technique, and the agreement between the two approaches is very good. Another advantage of the LTD approach is the smooth massless limit, thanks to the mapping being constructed to deal with quasi-collinear configurations. It is therefore possible to put $m$ to 0 before integration and still get the correct result.
\section{CONCLUSION AND OUTLOOK}
Throughout this article, we generalized the FDU method to the massive case and applied it on the three-point scalar function. Based on the LTD theorem, this technique allows to bypass DREG by expressing real and virtual contributions in terms of phase-space integrals, and adding them after splitting the integration domain and applying a mapping between the two kinematics in each of the different regions. This allows a safe four-dimensional integration, without the need to deal with $\epsilon$ poles, which makes the numerical computations easier and faster. The FDU technique has been applied to physical processes such as $\gamma*\to q\overline{q}$ \cite{Sborlini:2016hat}, and can be extended to compute multi-leg multi-loop processes as explained in \cite{Sborlini:2016gbr}.
\section{ACKNOWLEDGMENT}
This research project has been done in collaboration with Germ\'an Rodrigo and Germ\'an Sborlini. This work is partially supported by the Spanish Government and EU ERDF funds (grants FPA2014-53631-C2-1-P and SEV-2014-0398) and by GV (PROMETEU II/2013/007).

\end{document}